\documentclass{svproc}
\usepackage[utf8]{inputenc}
\usepackage[T1]{fontenc}
\usepackage{amsmath}
\usepackage{graphicx}
\usepackage{booktabs}
\usepackage{tabularx}
\usepackage{multirow}
\usepackage{enumerate}
\usepackage{xcolor}
\usepackage[misc,geometry]{ifsym}

\newcommand{\citep}[1]{\cite{#1}}
\usepackage{url}

\begin{document}
\mainmatter

\newcommand{\tb}[1]{\textcolor{blue}{#1}}
\newcommand{\tr}[1]{\textcolor{red}{#1}}

\title{Quoting is not Citing:\\Disentangling Affiliation and Interaction on Twitter}
\titlerunning{Quoting is not Citing}
\author{Camille Roth${}^{\text{\Letter}}$ \and Jonathan St-Onge \and Katrin Herms}
\authorrunning{Camille Roth et al.}

\institute{Computational Social Science team, Centre Marc Bloch, CNRS / Humboldt Universität, Friedrichstr. 191, D-10117 Berlin, Germany
\medskip\\
\email{\{roth, jonathan.st-onge, katrin.herms\}@cmb.hu-berlin.de}}

\maketitle

\begin{abstract}Interaction networks are generally much less homophilic than affiliation networks, accommodating for many more cross-cutting links. By statistically assigning a political valence to users from their network-level affiliation patterns, and by further contrasting interaction and affiliation (quotes and retweets) within specific discursive events, namely quote trees, we describe a variety of cross-cutting patterns which significantly nuance the traditional ``echo chamber'' narrative. 
\keywords{discussion trees; cross-cutting interaction; Twitter}
\end{abstract}

The socio-semantic assortativity of online networks is now a classical result: at the macro level, social clusters are often semantically homogeneous, exhibiting for instance similar political leanings \citep{adam-poli,himelboim2013birds}; at the micro level of users, links form more frequently between semantically similar dyads \citep{schi-folk,cinelli-2021-the-echo}. These observations depend nonetheless heavily on topics \citep{barbera-2015-twe,Garimella:2016:QCS:2835776.2835792}, and on link types: in particular, affiliation links generally configure networks where homophily is much stronger than with interaction links \citep{roth-soci}. On Twitter, this dichotomy separates subscriptions (followers) and (dry) retweets, from mentions and replies, whereby the latter are more cross-cutting than the former \citep{cono-poli,liet-when}. 
By focusing on quote cascades on Twitter \hbox{i.e.,} rather short-lived discursive events featuring in the same instance both link types (namely, quotes and retweets), we aim to examine the simultaneous manifestations of the affiliation/interaction dichotomy, which is normally studied in a separate or aggregate manner.  
Tweet cascades, or retweet trees, have long been studied from a diffusion perspective. Such trees are heterogeneous structurally \cite{kwak-what} and generatively, for instance alternating broad and deep propagation dynamics \cite{goel-2016-the-structural}; their formation speed and their range depends on content type, such as true \hbox{vs.} false news \cite{vosoughi2018spread}.  Quote tweets, or tweets with comments, appeared more recently (2015) even though they remind of the original conversational use of retweets \citep{boyd2010tweet} before becoming a proper tool on the platform. 
Research on quotes is still relatively sparse but confirms they are instrumental in (possibly antagonistic) conversation rather than propagation \cite{garimella2016quote,guerra2017antagonism}. 

A related strand of research has questioned whether online public spaces stimulate the development of like-minded groups or foster the exposure to diverse content \citep{lev-on-2009-happy,barbera2020social}. For one, beyond a commonly observed right-/left-wing biclustering, aggregate Twitter networks exhibit a mix of supportive and oppositional relationships \citep{vaccari-2016-echo} and a certain asymmetry whereby mainstream content receives much more attention from so-called ``counter-publics'' than the other way around \citep{kaiser-2017-alliance} --- all of which hints at a diversity of attitudes towards cross-cutting content and interactions. As we shall see, quote cascades on Twitter gather ephemeral publics that are generally local, in terms of time and of participants. By examining the structure of cross-cutting participation in quote trees, we also aim to contribute to study how local online arenas of a certain political orientation attract participants affiliated with diverse political orientations. In this regard, a series of recent results go against the grain of the traditional ``echo chamber'' narrative: users appear to engage heavily with content affiliated with an opposite camp, such as commenting on YouTube videos of some opposing channel \citep{wu-2021-cross-partisan} or posting messages on a Reddit thread of some opposing ``subreddit'' \citep{morales2021no}; more precisely, there exists a continuum of roles where users are diversely embedded in bipartisan networks \hbox{i.e.,} are at the interface between users of opposing political affiliations, or not \citep{garimella-2018-political}.

In a nutshell, we aim to describe the local and largely ephemeral quote tree structure in regard to the political valence both of the original content and of the users who further participate in trees in various ways; the valence is itself computed from a network observed on a much wider temporal and topological scale, thus serving as a basemap. This enables us to distinguish a variety of cross-cutting interaction patterns and roles.

\section*{Empirical data}

\paragraph{Perimeter and collection.} 
Over the whole year of 2020, we collected all publications by French-speaking Twitter users belonging to a perimeter based on the 2019 European Parliament elections. We had previously collected all tweets containing at least one hashtag among \{\#EU2019,  \#ElectionsEuropeennes2019, \hbox{\#Europeennes2019}, \#EP2019, \#Européennes2019, \#electionsue19,  \#CetteFoisJeVote\} between one month before and one month after the vote (April 26-June 28, 2019), focusing on users active in French (\hbox{i.e.,} publishing at least 15\% of tweets in that language). We further required users to have published at least 5 tweets over this period (\emph{minimum activity}) and be above the median number of 195 followers (\emph{minimum visibility}), which reduced the number of users from 39,938 to 15,919, of which 14,102 were still active in January 2020, and 13,074 in December 2020; reflecting a relatively low attrition rate given the initial focus on 2019 elections. Casual manual examination of this perimeter indicates that there are very few bots and that most well-known news sources or political figures have been included, thus suggesting that it represents a meaningful part of the politics-related French online Twitter space.

\begin{figure}[!t]\centering
\includegraphics[width=\linewidth]{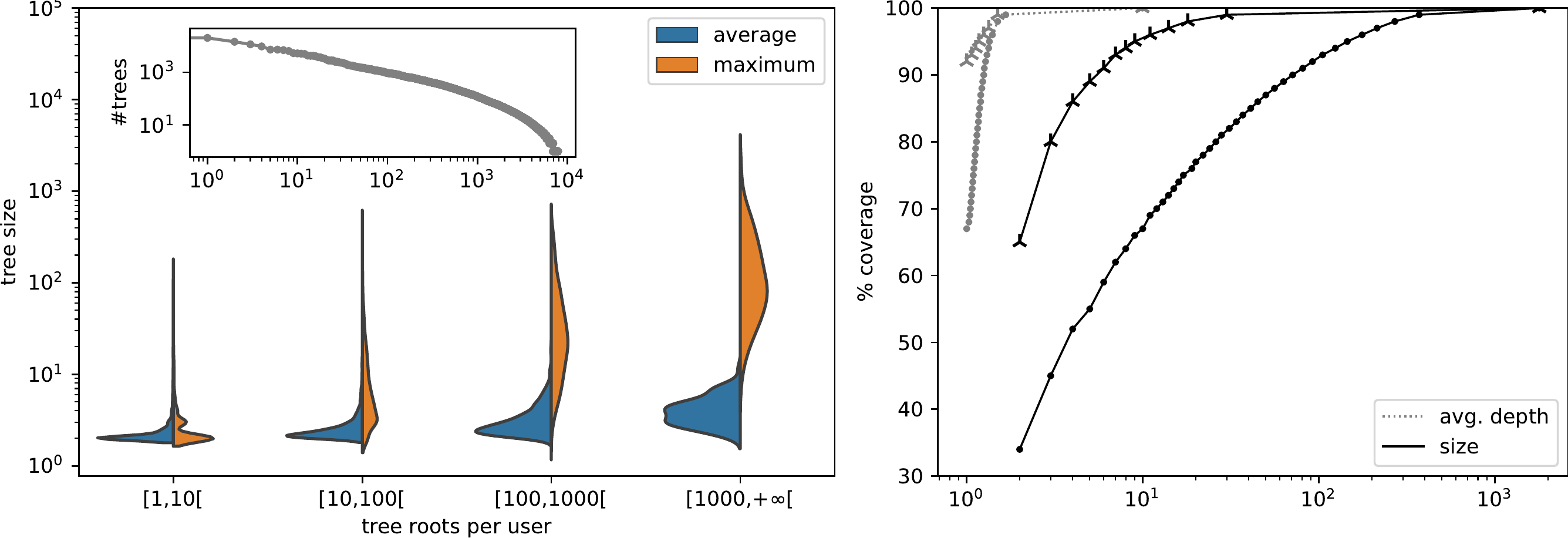}
\caption{\label{fig:urtweetsperuser}{\em Left:} in the inset, the number of tree roots per user follows a heterogeneous distribution; violin plots show the distribution of tree size (average or maximum) for users having generated a certain number of trees (few, more, many, very many).
\label{fig:treesizevsUT}{\em Right:} coverage of the dataset attained by focusing on trees up to a certain size (solid black line) or average depth (dotted gray line), in terms of the proportion of covered nodes (\hbox{i.e.,} quotes; represented as dots) or trees (represented as trihedrons).}
\end{figure}

\paragraph{Tree size and depth.}
We then build all non-trivial quote trees stemming from a initial tweet, or \emph{root tweet}, published in 2020. More precisely, we consider recursive cascades of quotes, restricted by construction to quotes from perimeter users, while excluding quotes where a user quotes themselves, and comprising at least one quote.
The dataset features 1.13m trees generated by 12,462 unique users \hbox{i.e.,} about 90 trees per active user, following a usual heterogeneous distribution (inset of Fig.~\ref{fig:urtweetsperuser}-left). Top users are unsurprisingly accounts of media and political figures generating in excess of 10k trees over the whole year \hbox{i.e.,} dozens a day. Besides,  trees of more prolific users are generally larger on average and among the largest ones (violin plots on Fig.~\ref{fig:treesizevsUT}).
Tree size also follows a heterogeneous law whereby 75\% of all trees are of size 2 or 3, 90\% of size 6 or less, and only 1\% are larger than 30 nodes, as shown on Fig.~\ref{fig:treesizevsUT}-right. By definition, larger trees gather more quotes and thus represent a larger portion of the dataset in relative terms. To keep the focus on quotes and avoid an over-representation of relatively trivial and very small trees in the subsequent computations, we rather consider the coverage of the dataset in terms of tree nodes. This leads us to define thresholds of {\em small}, {\em medium} or {\em large} trees by considering respectively a coverage of 75\% of all nodes (trees containing up to 17 nodes), 90\% or less (up to 71 nodes), and the last decile (remaining trees up to a maximum of 1786 nodes).

\begin{figure}[t]\centering
\includegraphics[height=3.15cm]{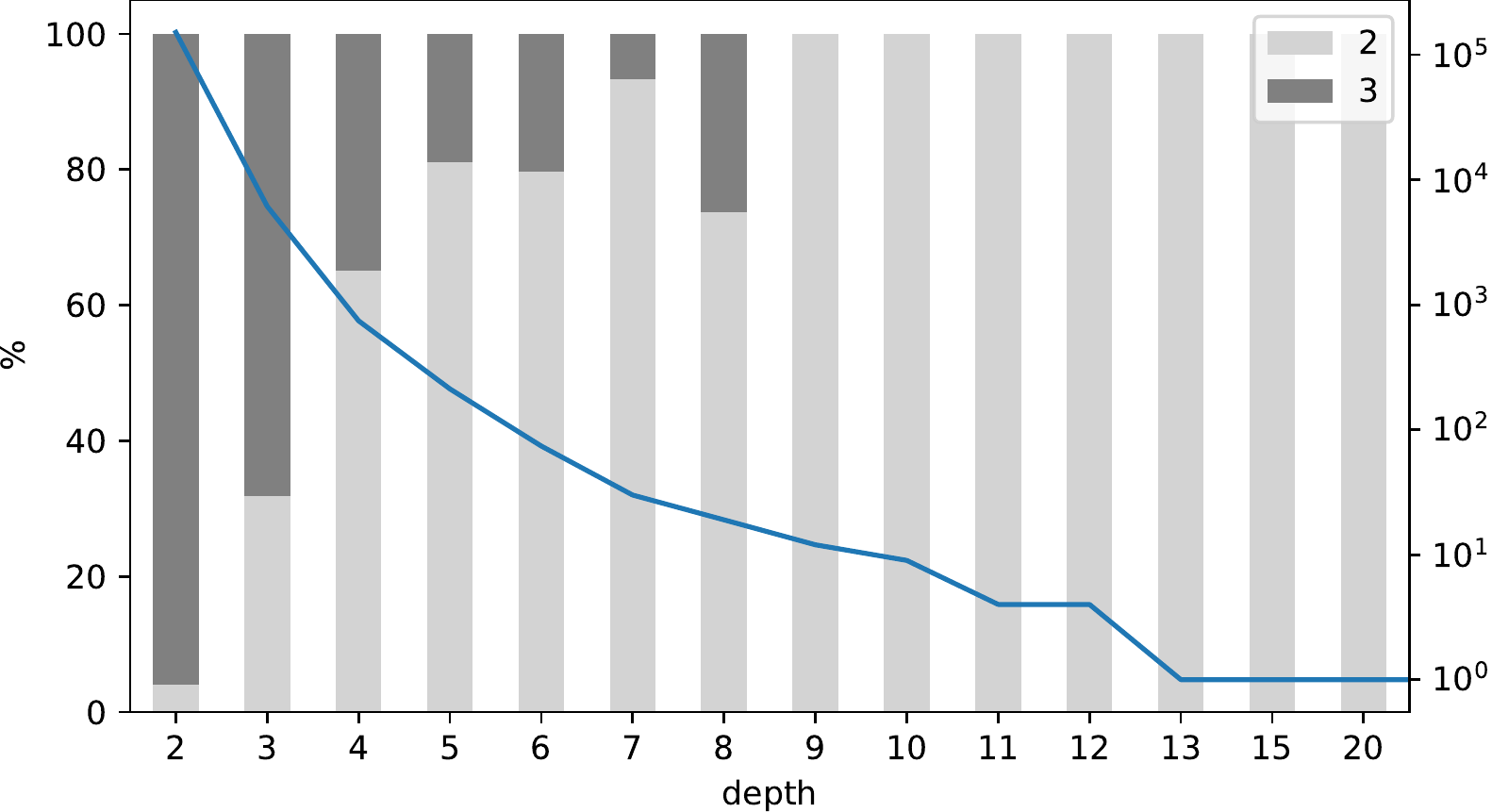}
\includegraphics[height=3.185cm]{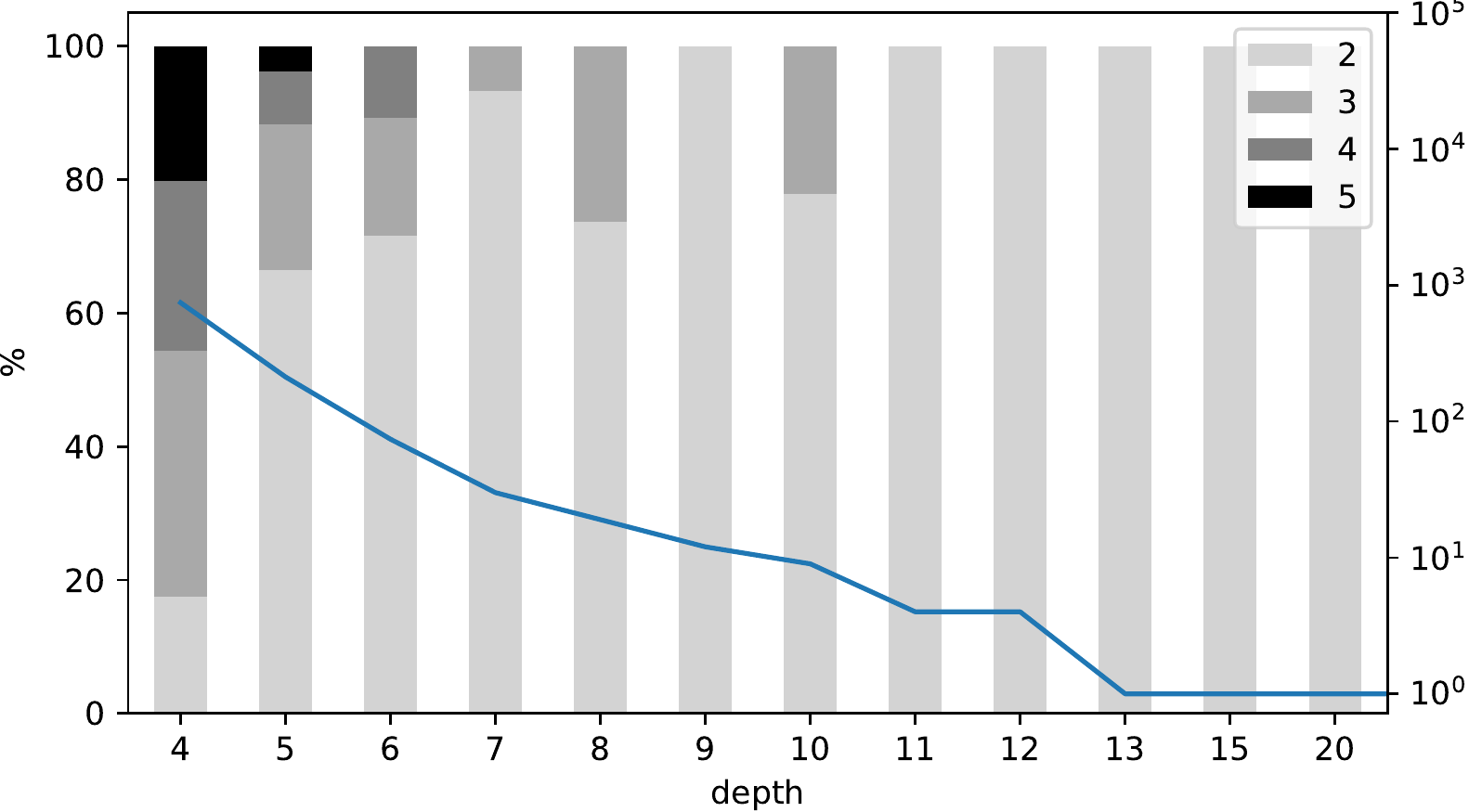}
\caption{\label{fig:uniquequoters}{\em Left:} number of unique quoters in subchains of size 3, which equals two when there is redundancy (A-B-A) and three otherwise (A-B-C), since we ignore self-quotes (A-A-*). {\em Right:} number of unique quoters in subchains of size 5.}
\end{figure}

\newcommand{\avgd}{\langle d\rangle}
The average depth of trees, denoted as $\avgd$ and computed as the average distance from the root tweet over all nodes, is generally small, with more than 90\% of trees with a $\avgd$ of 1 (Fig.~\ref{fig:treesizevsUT}-right), indicating the absence of secondary quotes, or quotes of quotes. Less than 2\% of trees feature a $\avgd>1.5$ (majority of secondary quotes) and less than 3\% of nodes belong to such trees. On the whole, depth is a relatively rare phenomenon, as shown by the exponentially decreasing number of chains reaching a certain depth over all trees (solid line on \hbox{Fig.~\ref{fig:uniquequoters}}). Furthermore, deeper chains correspond to ping-pongs between two individuals (A-B-A-B...) rather than iterative quoting between distinct users (A-B-C-D...): to show this, we plot the number of distinct quoters in a chain, as a function of its maximal depth, focusing on terminal subchains of a given length $w$. In other words, we look at the composition of the $w$ last quoters of a chain of some depth. Histograms on Fig.~\ref{fig:uniquequoters} rely on $w=3$ and $w=5$ (results are similar for other window sizes $w$) and show a strongly increasing proportion of chains of only two distinct individuals (out of $3$ or $5$ possibilities) when going deeper in the tree. Such ping-pongs may correspond to a dialogical framing behavior where two users conflate the quote and reply functions. In any case, they represent a tiny portion of the data. 

\paragraph{Political valence of users.}

\newcommand{\IP}{\text{IP}}
We define the likely political position of users by estimating their so-called ``Ideal Point'' (IP), a technique first introduced to infer a unidimensional political valence of lawmakers from the set of bills they support \citep{poole2005spatial} and more recently applied on Twitter users based on the set of accounts they follow \citep{barbera-2015-bir}. This method relies on the manual attribution of a fixed valence to a small subset of bootstrap users,  or ``elites'', from which positions are computed for the whole dataset along affiliation links. We use here the set constructed by 
\citep{briatte-2015-recovering} comprising 2,013 elites of the French political realm. We then collected the follower set for all users of our dataset (as of January 2021). We were eventually able to compute the IP value of 9,815 users who follow at least 10 elites which provided enough information for the IP estimation. 

\begin{figure}[!t]
\centering
\includegraphics[width=.8\linewidth]{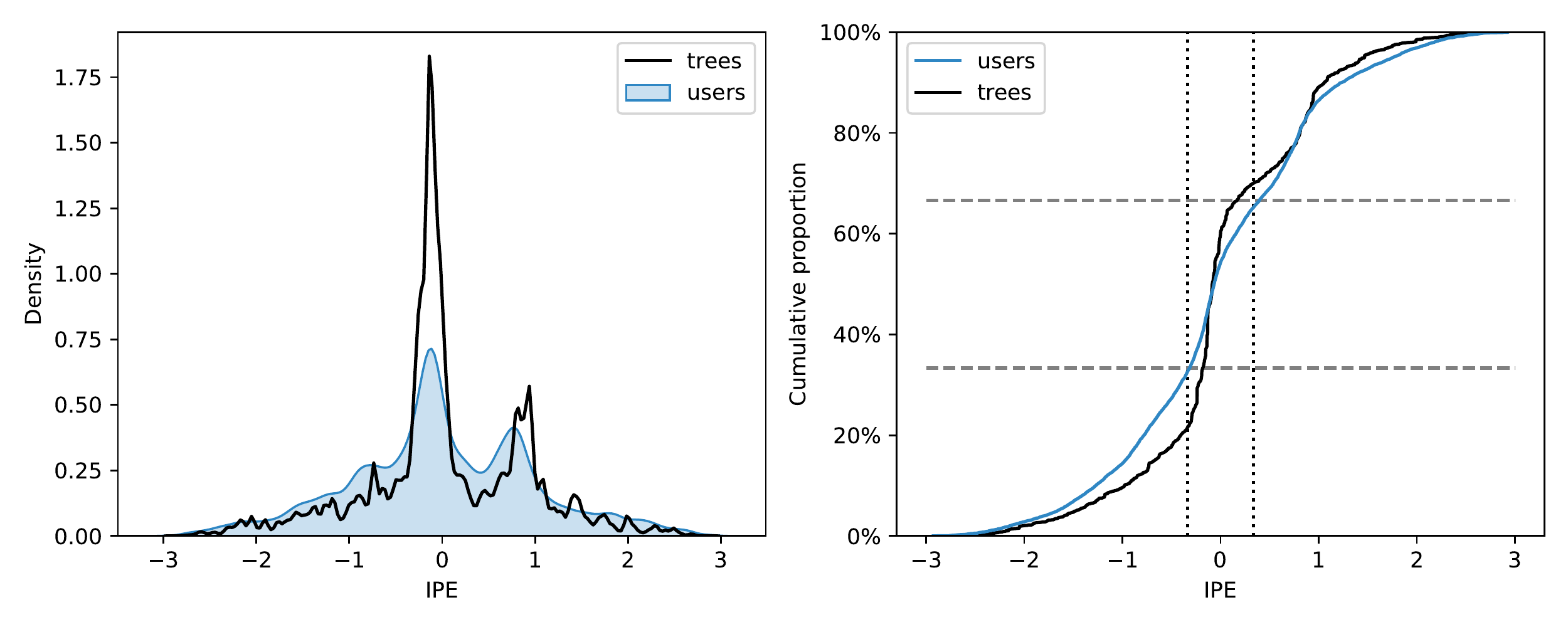}\vspace{-1em}
\caption{\label{fig:IPusers}{Distributions of IP values for corpus users and for tree root users, kernel density estimations (\emph{left}) and cumulative distributions (\emph{right}).}}
\end{figure}

We observe in Fig.~\ref{fig:IPusers} that IPs may roughly be broken down into three ranges gathering each a third of the density: markedly negative values where $\IP<-\frac{1}{3}$; somewhat central values around 0, $\IP\in[-\frac{1}{3},\frac{1}{3}]$; and markedly positive values, $\IP>\frac{1}{3}$. For users whose political affiliation is explicitly known, who are also well represented in our dataset which further confirms our good coverage of the political space, these three ranges match what is usually considered as left-wing, center and right-wing, respectively. For instance, all users who are explicitly members of PS (Parti Socialiste, left-wing) have an IP below 0 with an average around -1; while all members of LR (Les Républicains, right-wing) have an IP above 0, of average around +1. Without entering into a debate concerning the relevance of political labels based on unidimensional values, we deem IPs to be a sufficient proxy to characterize the relative political positions of users generating and participating in quote trees.

\section*{Tree structure and quoting behavior}

To describe quote trees in relation to the political valence of their root author, we now exclusively focus on the {699k} trees whose root tweet user has a known IP, denoted $\rho$. This makes about two thirds of all trees. Relative to user IPs, the distribution of $\rho$ over trees favors central and, to a lesser extent, right-tilted values (essentially close to $+1$). More precisely, half of tree roots stem from the third of users with a central IP, while about 20\% stem from users with a markedly negative IP (left), and 30\% with a markedly positive IP (right, with the same peak around $+1$). Upon casual examination the 30 top accounts generating the most trees, which are thus also larger, belong mainly to mainstream media organizations and, to {a lesser} extent, center-wing political figures.

\paragraph{\bf General features.}

\begin{figure*}[!t]\centering
\includegraphics[width=\linewidth]{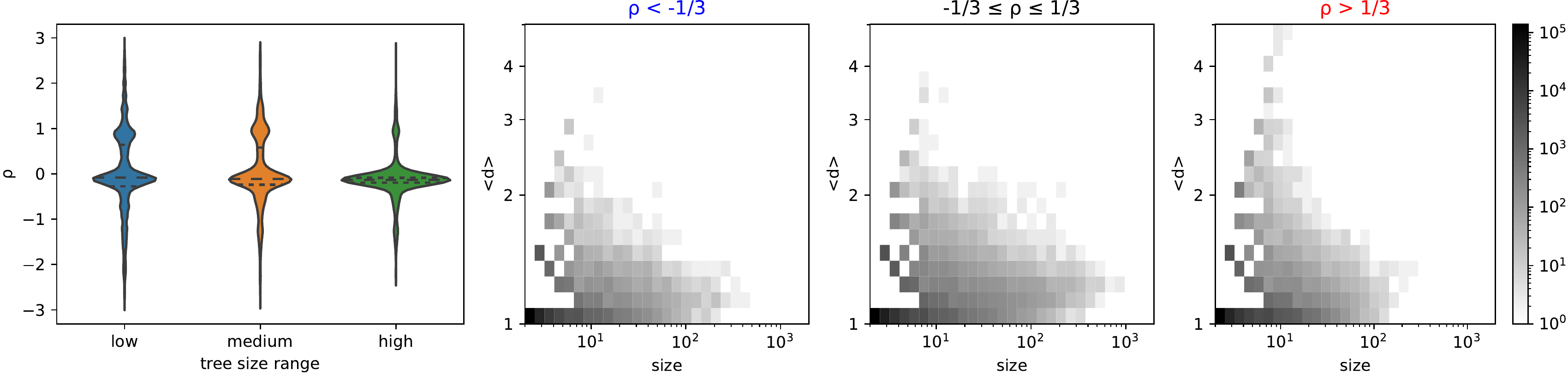}
\caption{\label{fig:treeshapeIP}{{\em Left:} Root user IP $\rho$ as a function of tree size range. {\em Right panels:} heatmap of trees having a certain average depth $\avgd$ and size, for each of the three IP ranges.}}
\end{figure*}

We first consider the relationship between size, average depth $\avgd$, and root IP $\rho$. Results are summarized in Fig.~\ref{fig:treeshapeIP}. The left panel shows the distribution of $\rho$ for the three tree size ranges. The largest trees are more often generated by central IP users. The right panels show heat maps for each IP range and three interesting areas, in decreasing order of density: (1) both shallow and small- and medium-sized trees, by far the most frequent ones over the whole spectrum, (2) medium- to large-sized trees and moderately deep, which seem to be more often generated by central IP users, (3) small yet deep trees, whose root is more often made of IP-positive users when focusing on the 1\% deepest trees (indicative of a narrow reach with strong tendency to long chains).

\newcommand{\avgR}{\langle R\rangle}
\newcommand{\avgQ}{\langle Q\rangle}

\paragraph{\bf First-order layer.}
\begin{figure}[b!]\centering
\includegraphics[width=\linewidth]{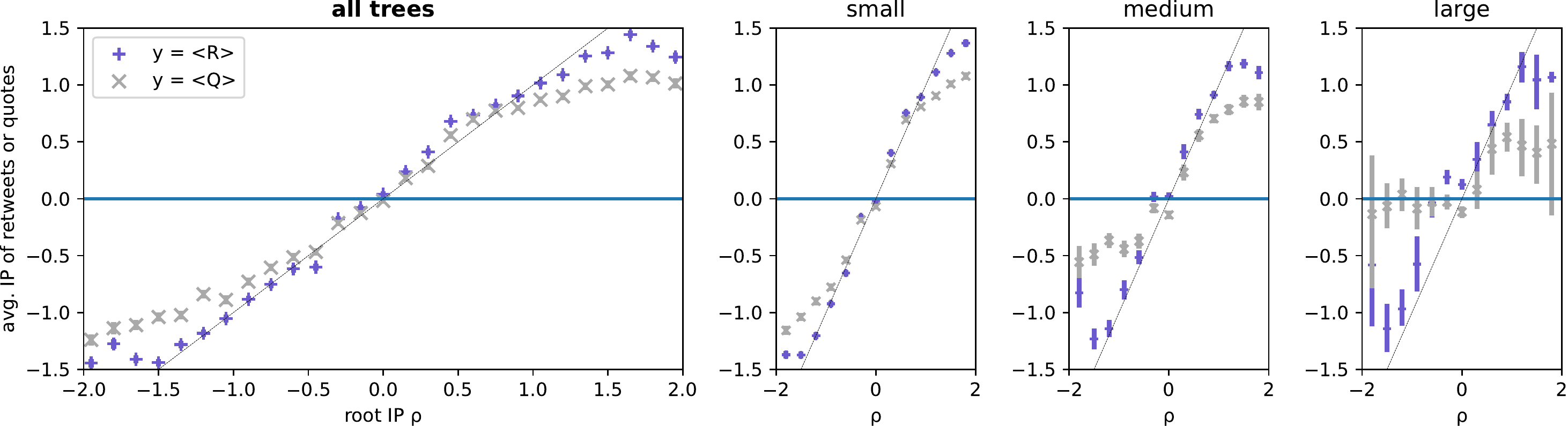}
\caption{\label{fig:bimodalityvsquotes}Average IP of retweeters $\avgR$ or quoters $\avgQ$ of a root tweet from a user of IP~$\rho$. {\em Right panels:} breakdown by tree size.}
\end{figure}

Based on the above, we contend that focusing on the two first layers (\hbox{i.e.,} primary and secondary quotes) captures most of the content framing behavior. 
We examine the average IP of the first layer of quotes, denoted as $\avgQ$, with respect to the root user's IP value $\rho$. We also consider $\avgR$, the average IP of so-called ``dry'' retweets of the  root tweet \hbox{i.e.,} without quoting, which we deem a proxy of its political position: retweets indeed correspond to the audience of users who plainly forward with no further framing. On the whole, we observe on Fig.~\ref{fig:bimodalityvsquotes} that $\avgR$ generally follows $\rho$. Average IP values of quoting users, by contrast, tend to diverge from both $\avgR$ and $\rho$ when $\rho$ is not central, all the more for large trees as indicated on the three small panels. 
In other words, tweets at the root of large trees generate quotes, or framing instances, from the whole political spectrum, irrespective of the position of their retweeting audience; while smaller trees exhibit a narrower spectrum of quoting reactions, closer to $\rho$ and thus both the root and $\avgR$.
Note that the standard deviations of $\avgR$ and $\avgQ$, not shown here, are relatively constant across these spectrums ---around $0.65$--- indicating some amount of variability around each average.

\begin{figure}[t!]\centering
\includegraphics[width=\linewidth]{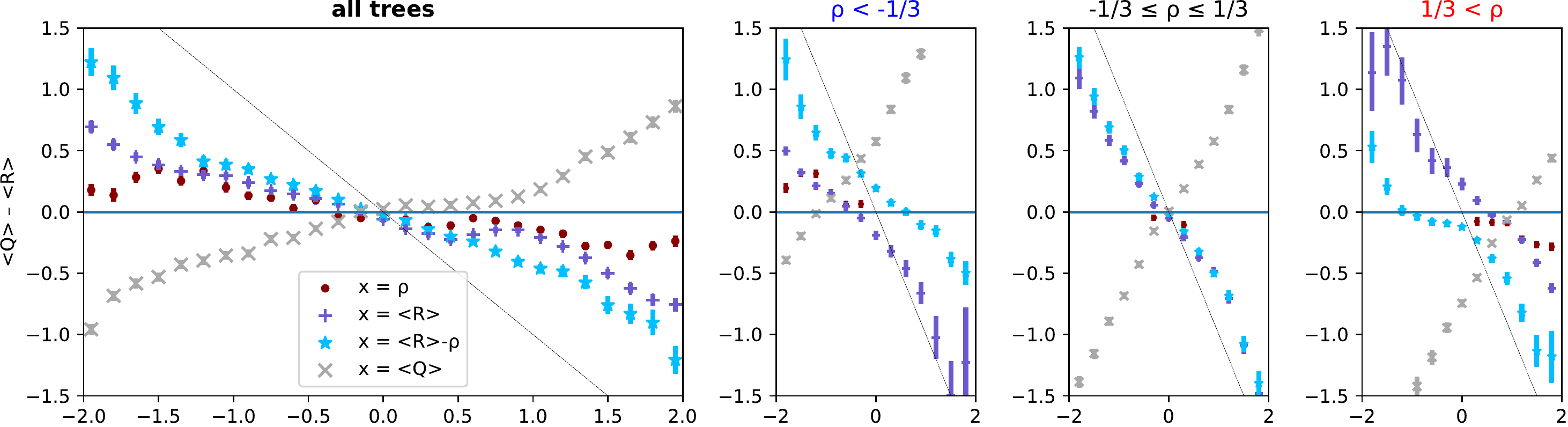}
\caption{\label{fig:deltaQR}{Discrepancy between the average IP value of quotes vs. retweets. {\em Right panels:} breakdown by IP category of the root.}}
\end{figure}

Figure~\ref{fig:deltaQR} characterizes further this divergence between quotes and retweets. We compare $\avgQ-\avgR$ with: the root IP $\rho$ (i.e., the difference between the two curves of the previous figure), the average IP of retweeters $\avgR$, as well as the \emph{offset} between retweets and the root IP, $\avgR-\rho$. This last quantity indicates how far the retweeting population of a root tweet is from the (constant) IP of the root user. We observe first that the divergence $\avgQ-\avgR$ goes, on average again, in a direction opposed to the IP value considered on the x-axis: be it $\rho$, $\avgR$ or $\avgR-\rho$. For instance, divergences are increasingly negative for positive root, retweet and \emph{offset} IP; and the other way around. They also remain under the $y=-x$ curve: the magnitude of this ``backlash'' is thus smaller than the initial shift from the center (IP=0). Put simply, if a root tweet is tweeted or retweeted by the left, \emph{on average}, it is still going to be quoted, \emph{on average}, by the left, but less so. Second, magnitudes of $\avgQ-\avgR$ are larger when compared with $\avgR$, and even more with $\avgR-\rho$, than with $\rho$: they are stronger when a root tweet attracts retweets from non-central users as well as from users off the ``baseline'' IP of the root. The small middle panel in Fig.~\ref{fig:deltaQR} is illustrative in this regard: it focuses on trees produced by central users ($-\frac{1}{3}\leq\rho\leq\frac{1}{3}$) whose discrepancy $\avgQ-\avgR$ is in aggregate close to 0 (as per Fig.~\ref{fig:bimodalityvsquotes}). Yet, even for these root tweets from central users, $\avgQ-\avgR$ grows as $\avgR$ or $\avgR-\rho$ diverge from 0.

To summarize, first-layer quotes diverge more from retweets in larger trees and when root tweet users are non-central, and even more so when average retweeters are non-central or unusually off the root IP. Here again, standard deviations stay around $0.65$ for all curves, indicating nonetheless a varied constellation of situations. Assuming that retweets are, on average, rather concentrated around the same IP as the root tweet user, these observations configure an instantaneous, low-level dynamics at the level of individual trees where quotes are all the more off the ``baseline'' of the retweeting population as this population is off the root user value. This makes it possible to hypothesize that quotes feature local counter-publics of users who come from a distinct set of IP positions to intervene to frame the original content. We thus turn to users.

\begin{figure}[t!]\centering
\includegraphics[width=\linewidth]{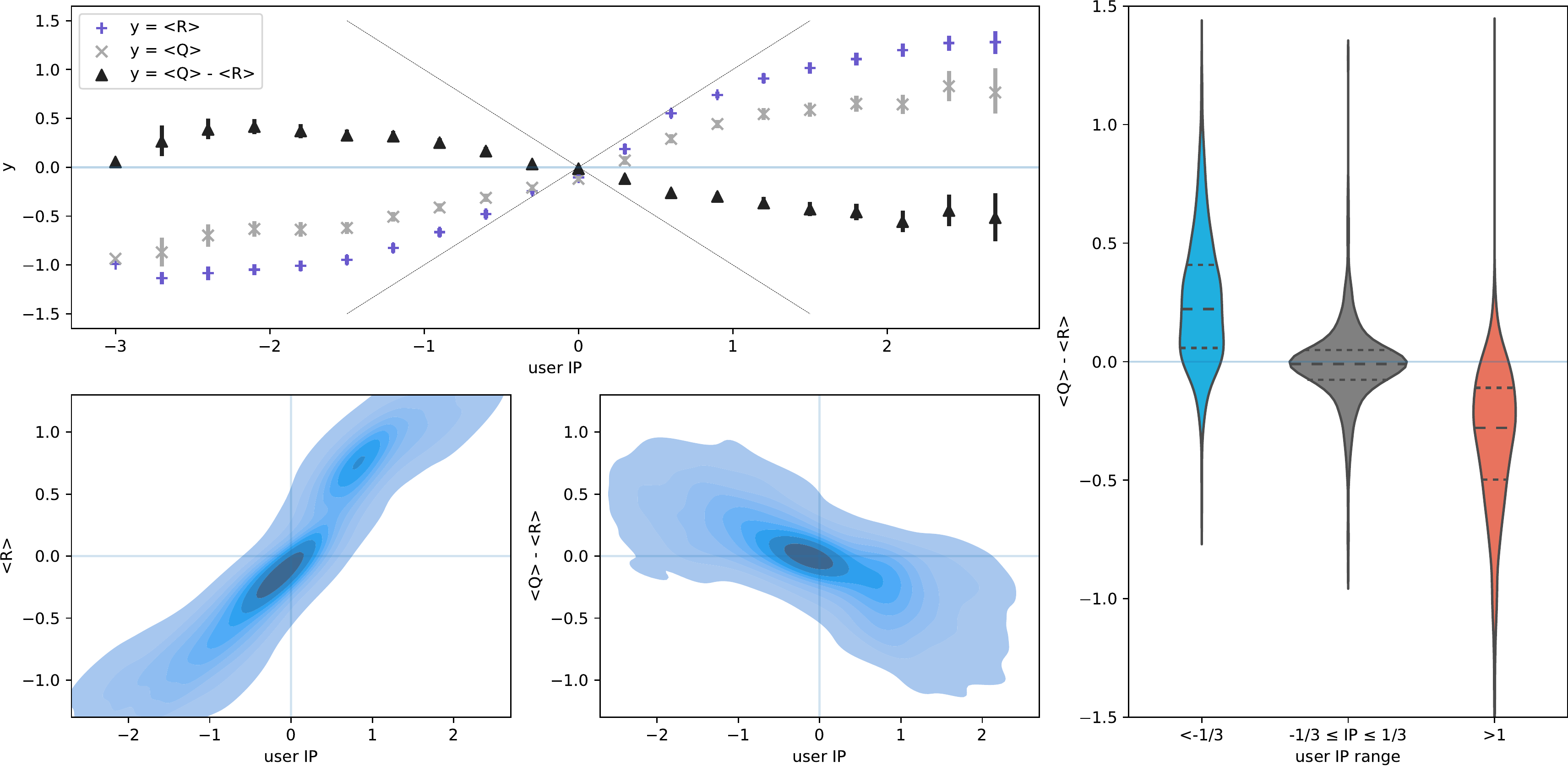}
\caption{\label{fig:deltaQRusers}{User-centric discrepancy of quote vs. retweet behavior.}}
\end{figure}

\paragraph{\bf User-centric patterns.} Several of these tree-centric observations hold from a user-centric perspective. Figure~\ref{fig:deltaQRusers} confirms that users retweet roots roughly along their own IP on average, albeit less so for extreme users (which may probably partly explained by artefactual reasons where \hbox{e.g.,} users of IP $>2$ do not have much content to retweet on their right). Quotes are however more diverse and as a result the divergence $\avgQ-\avgR$ is not flat and higher in absolute values for non-central users. The bottom left heat map illustrates the former point, the bottom central heat map the latter one. 

Moreover, the second heat map underlines a higher spread of $\avgQ-\avgR$ for non-central users, some of them exhibiting an average divergence close to 0 (quoting on the same material they would retweet), others exhibiting a high average divergence.
This hints not only at the existence of various roles, but at the higher spread of these roles further from the center. Violin plots in Fig.~\ref{fig:deltaQRusers}-right support this interpretation: the top (respectively bottom) quartile for users with a negative (respectively positive) IP is above (respectively below) 0.  Albeit beyond the scope of this paper, it would be most interesting to examine qualitatively who these users are, both from the content they publish and from interviews, to contrast their interest in participating in an online public space.

\begin{figure}[t!]\centering
\includegraphics[width=.999\linewidth]{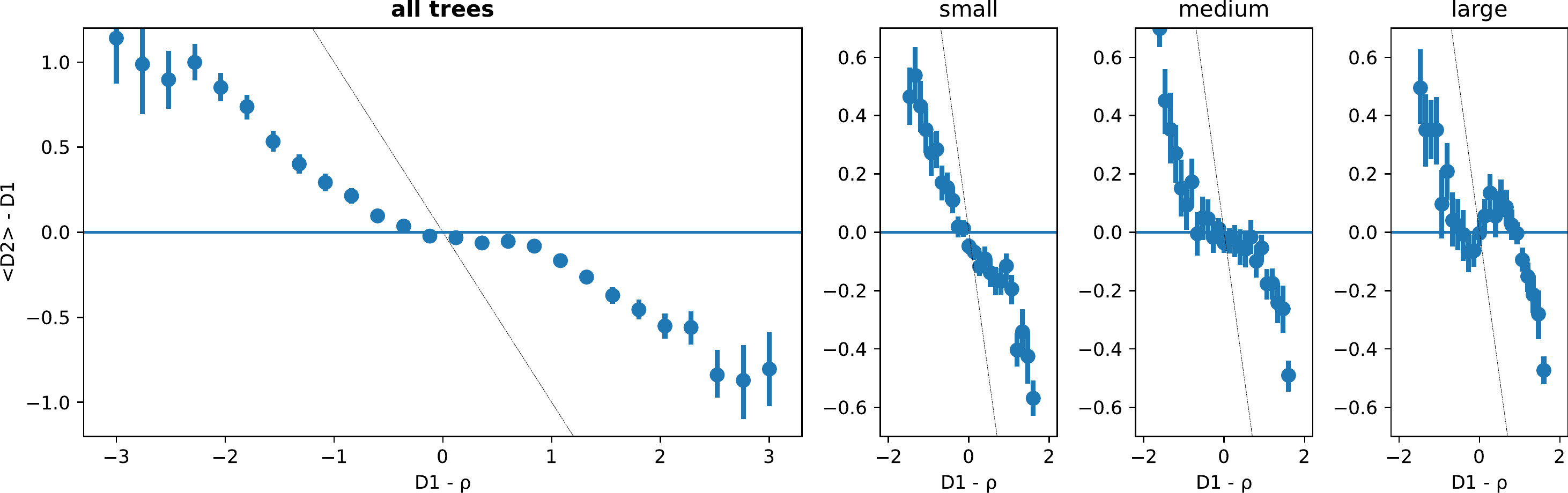}
\caption{\label{fig:depth2}Comparison of the discrepancy between a primary quoter's IP $D1$ and the root IP $\rho$ {\em (x-axis)} and the average discrepancy between IPs of secondary quoters and that of their immediate parent $D1$ {\em (y-axis)}. {\em Panels:} breakdown by tree size ranges.}
\end{figure}

\paragraph{\bf Quotes of quotes: toward the deeper layer.}
While primary quotes tend to go against the polarity of the initial root tweet (all the more for non-central roots), in relative terms and all other things being equal, it is unclear whether these dynamics persist deeper in the tree and, for one, whether secondary quotes are made by users whose IP is more aligned with that of primary quoters or not (toward, or away from, the root tweet user). To shed light on this issue, we simply compare the discrepancy between a primary quoter's and the root's IPs, $D1-\rho$, with the discrepancy between that primary quoter and their secondary quoters, $\langle D2\rangle-D1$. We observe in Fig.~\ref{fig:depth2} that secondary quotes tend to turn the tide \hbox{i.e.,} they stem from users yet again closer to the root, all the more when the primary quoter's IP diverges from the root. In other words, if a quoter is more to the left than the root, a secondary quoter is going to be more to the right than the quoter, in a sort of back-and-forth movement. 

The amplitude of this movement is however smaller at the second level – it is as if the shift of secondary quotes was damped: the IP value of the second quoter is, on average, less far from the first quoter, than the first quoter is from the root. 
Interestingly, the direction of this movement is non-monotonous for the largest trees, where second quoters are further in the \emph{same} direction as the first quoters for small discrepancies ($D1-\rho)$, while this trend gets reverted for larger discrepancies (rightmost panel of Fig.~\ref{fig:depth2}). Put differently, for root tweets generating the largest numbers of quotes, there appears to be two types of quotes: those originating from quoters close to the root IP and which further attract second quoters roughly of the same polarity, and those originating from quoters that are farther and attracting second quoters in the opposite direction.

\paragraph{\bf Qualitative homogeneity of some framing practices.}

\begin{figure}[!t]
\centering
\includegraphics[width=.862\linewidth]{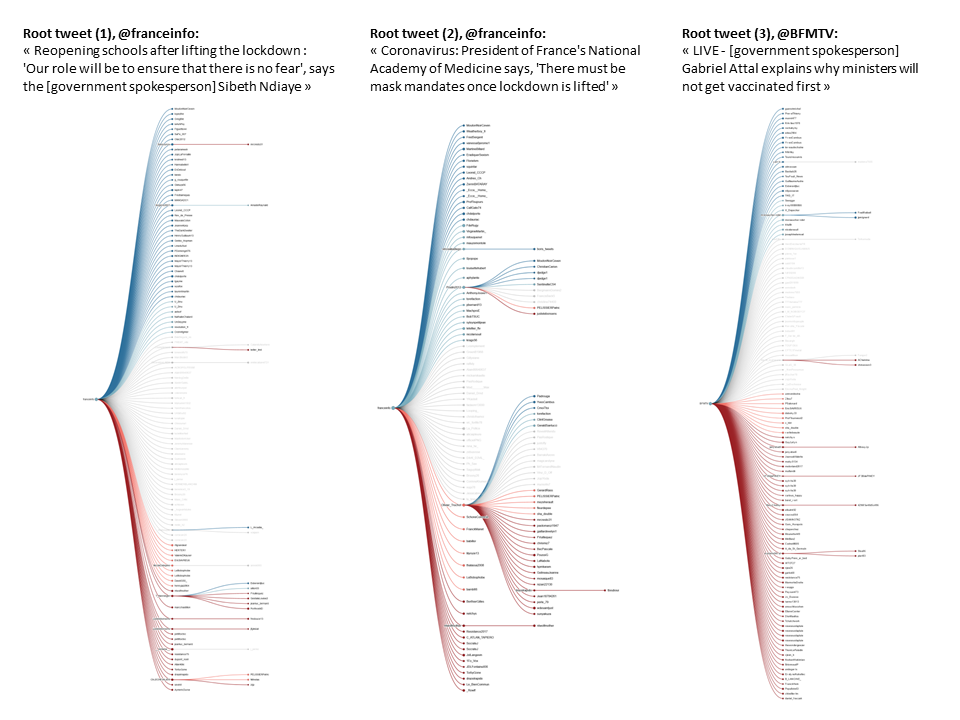}\vspace{-1.14em}
\caption{\label{fig:illustration}Structure of 3 illustrative trees. Nodes are colored in blue when users have an IP<-$\frac{1}{3}$, black for central IP values $\in[-\frac{1}{3},\frac{1}{3}]$, red for IP>$\frac{1}{3}$. 
}
\end{figure}

We finally qualitatively illustrate one of our findings on the behavior of $\avgQ-\avgR$ by studying in more detail a handful of trees related to the above-mentioned example on the small middle panel in Fig.~\ref{fig:deltaQR} \hbox{i.e.,} with a central $\rho$ close to 0. We focus on large trees, to ensure a meaningful qualitative analysis, and on keywords related to the main political measures to curb the Covid-19 crisis in France (admittedly one of the most debated issues in 2020), to ensure comparability among trees. To exemplify each region of this graph, we arbitrarily select three trees whose $\avgR$ is respectively negative, close to zero, and positive. They respectively deal with (1) lockdown lifting ($\avgR=-.63$, $\avgQ=-.48$), (2) mask mandates ($\avgR=.11$, $\avgQ=-.22$), and (3) vaccination ($\avgR=.80$, $\avgQ=.30$); see Fig.~\ref{fig:illustration}.
Overall, we expectedly observe participation from the whole spectrum.

\newcommand{\BU}{\tb{\textbf{<}}}
\newcommand{\CU}{$\mathbf{\sim}$}
\newcommand{\RU}{\tr{\textbf{>}}}
\newcommand{\B}[1]{\tb{#1}}
\newcommand{\Z}[1]{\textbf{#1}}
\newcommand{\R}[1]{\tr{#1}}
\newcommand{\BX}[1]{\tiny\tb{\em#1}}
\newcommand{\CX}[1]{\tiny{\em#1}}
\newcommand{\RX}[1]{\tiny\tr{\em#1}}
\newcommand{\vps}{}

\begin{table}[!h]\scriptsize{\sf
\begin{tabularx}{\linewidth}{>{\bf}lp{7cm}|rrrr|rrrr|rrrr}
\toprule
&Frame category&\multicolumn{4}{l|}{Tree 1}&\multicolumn{4}{l|}{Tree 2}&\multicolumn{4}{l}{Tree 3}\\
&\tiny(nuances in parenthesis for subtopics specific to tree [1], [2] or [3])&
\tiny total&\BU&\CU&\RU&\tiny total&\BU&\CU&\RU&\tiny total&\BU&\CU&\RU
\\\midrule
A& \multirow{2}{7cm}{Call for responsibility (to [1] ensure a secure school opening, [2] improve crisis management, [3] test the vaccine first)}
&	\Z{19}&	\B{13}&	2&	\R{4}
&	\Z{26}&	\B{10}&	5&	\R{11}
&	\Z{17}&	\B{5}&	1&	\R{11}
\\
&
&\tiny\%&\BX{30}&\CX{17}&\RX{13}
&\tiny\%&\BX{31}&\CX{21}&\RX{31}
&\tiny\%&\BX{23}&\CX{7}&\RX{17}
\\
B&	Allegation of incompetent political communication
&	\Z{31}&	\B{19}&	1&	\R{11}
&	\Z{31}&	\B{17}&	5&	\R{9}
&	\Z{17}&	\B{7}&	4&	\R{6}\\
&
&\tiny\%&\BX{43}&\CX{8}&\RX{37}
&\tiny\%&\BX{53}&\CX{21}&\RX{25}
&\tiny\%&\BX{32}&\CX{29}&\RX{9}
\\
C&	Allegation of malice: ``officials act against us, the people''
&	\Z{14}&	\B{8}&	2&	\R{4}
&	\Z{33}&	\B{13}&	8&	\R{12}
&	\Z{36}&	\B{7}&	4&	\R{25}\\
&
&\tiny\%&\BX{18}&\CX{17}&\RX{13}
&\tiny\%&\BX{41}&\CX{33}&\RX{33}
&\tiny\%&\BX{32}&\CX{29}&\RX{38}
\\
D&	\multirow{2}{7cm}{Argumentation (around fear and risk [1 \& 3] or solidarity and utility [2])}
&	\Z{20}&	\B{9}&	5&	\R{6}
&	 \Z{9}&	\B{3}&	3&	\R{3}
&	\Z{17}&	\B{5}&	2&	\R{10}\\
&
&\tiny\%&\BX{20}&\CX{42}&\RX{20}
&\tiny\%&\BX{9}&\CX{13}&\RX{8}
&\tiny\%&\BX{23}&\CX{14}&\RX{15}
\\
E&\multirow{2}{7cm}{Protest (against sending children to school [1], wearing a mask [2], getting vaccinated [3])}
&	\Z{2}&	\B{2}&	0&	\R{0}
&	\Z{4}&	\B{1}&	2&	\R{1}
&	\Z{4}&	\B{1}&	0&	\R{3}\\
&
&\tiny\%&\BX{5}&\CX{0}&\RX{0}
&\tiny\%&\BX{3}&\CX{8}&\RX{3}
&\tiny\%&\BX{5}&\CX{0}&\RX{5}
\\
F&	Insults and mockery against leaders
&	\Z{18}&	\B{7}&	2&	\R{9}
&	\Z{13}&	\B{5}&	2&	\R{6}
&	\Z{15}&	\B{2}&	0&	\R{13}\\
&
&\tiny\%&\BX{16}&\CX{17}&\RX{30}
&\tiny\%&\BX{16}&\CX{8}&\RX{17}
&\tiny\%&\BX{9}&\CX{0}&\RX{20}
\\
G&	Emotional exclamations
&	\Z{6}&	\B{3}&	1&	\R{2}
&	\Z{4}&	\B{1}&	0&	\R{3}
&	\Z{14}&	\B{2}&	4&	\R{8}\\
&
&\tiny\%&\BX{7}&\CX{8}&\RX{7}
&\tiny\%&\BX{3}&\CX{0}&\RX{8}
&\tiny\%&\BX{9}&\CX{29}&\RX{12}
\\
H&	Other
&	\Z{2}&	\B{1}&	0&	\R{1}
&	\Z{14}&	\B{3}&	5&	\R{6}
&	\Z{12}&	\B{3}&	2&	\R{7}\\
&
&\tiny\%&\BX{2}&\CX{0}&\RX{3}
&\tiny\%&\BX{9}&\CX{21}&\RX{17}
&\tiny\%&\BX{14}&\CX{14}&\RX{11}
\\\bottomrule
\end{tabularx}}
\caption{\label{tab:framegrams}%
Number of quotes featuring a given frame category, 
Frame categories for all trees (with ) and counts per tree, broken down by user valence (``\tb{\textbf{<}}'' for IP<-$\frac{1}{3}$, ``$\sim$'' central IP, ``\tr{\textbf{>}}'' IP>$\frac{1}{3}$). Percentages indicate the proportions of quotes of a given color that mention a given frame. Note: quotes using multiple frames appear several times.}
\end{table}

We specifically compare quotes from cross-cutting users with those from non-cross-cutting users \hbox{i.e.,} users who intervene on roots that are primarily retweeted by users from the opposite \hbox{vs.} the same side. As said before, quotes are framing operations and we naturally use the notion of ``frames'' to qualitatively detail their nature. A frame is defined as a rhetorical device to recontextualize root tweet issues through the lens of a certain perspective, including normative judgments \citep{entman-1993-framing}. We build frame categories using an inductive handcoding approach typical of the ``grounded theory'' \citep{glas-disc}, looking for semantic similarities among quotes of a given tree. We then grouped similar claims and detected 7 frame categories for each tree, which are also quite recurrent across trees, plus an eighth category, ``other'', which regroups rare and isolated frames. 

Most frames aim at criticizing officials and their abilities to curb the crisis. Categories differ essentially in the form of that criticism: ranging from concrete expectations (frame A), via expressed mistrust related to incompetent communication (frame B), to allegations of selfishness and malice (frame C), plain protest (frame E) or even insults and mockery (frame F). Table~\ref{tab:framegrams} shows the breakdown of frame categories for each of the three trees and each user political valence/color.
We found that all colors are generally present in all frames. Some categories are used predominantly by a specific color: for instance, frame B (incompetence) is on average more often used by blue (43\%) than red (24\%) or black (19\%) quotes; as is, to a lesser extent, frame A (call for responsibility). By contrast, frame F (insults/mockery) tends to appear more in red (22\%) than blue (14\%) and black quotes (8\%). Interestingly, some frames are balanced, such as frames C (us/them) and D (argumentation). Keeping in mind the small size of this preliminary exploration, and even though there are remarkable variations in the use of some frames by some color, we nonetheless hypothesize that quote frames might obey a vertical dichotomy between ``us, the people'' and ``them, the officials'' as much as a moderate horizontal dichotomy between political camps  --- it is as if cross-cutting interventions fulfill a relatively similar rhetorical goal. 

\vspace{-.25em}
\section*{Concluding remarks}
We differentiated affiliation and interaction links on Twitter by focusing on a specific object featuring both link types: quote trees. We showed under which conditions these ephemeral discursive events may attract a diverse public eager to frame the initial information contained in the root tweet and coming from a more or less wide spectrum of estimated political valences. In particular, assuming that retweets reflect the ``baseline'' audience valence of a given root tweet, we observed that the public of quoters diverges all the more from baseline when the root tweet has a non-central valence and attracts a larger audience. Moreover, this back-and-forth movement persists in secondary quotes, albeit in an attenuated and non-monotonous manner. At first sight, these phenomena go against the ``echo chamber'' narrative, at least for larger ``chambers'' and trees.

Coming back to users, we nuanced this finding by exhibiting distinct user attitudes: while some users (especially non-central ones) quote root tweets of a distinct valence as the tweets they normally retweet, some users do not, reminiscing a behavior more akin to echo chambers. A casual yet in-depth qualitative exploration of just three trees further 
showed that both cross- and non-cross-cutting users nevertheless appear to rely on a small set of mildly shared frames. Put simply, cross-cutting interventions do not necessarily use cross-cutting frames. While shedding light on the formation and composition of counter-publics in reaction to content published in online social networks, our results hint at further research that would focus on specific regions of the figures presented in this paper, and qualify in more detail the position, behavior and claims of the corresponding users.

\bigskip
\scriptsize
\noindent{{\em Acknowledgments.} We are grateful to Telmo Menezes and Katharina Tittel for contributing to define the user perimeter and subsequently collect Twitter data. This work was supported by the "Socsemics" Consolidator grant from the European Research Council (ERC) under the European Union’s Horizon 2020 research and innovation program (grant agreement No. 772743)
}

\end{document}